\documentclass[12pt]{iopart}
\usepackage{amssymb}
\usepackage{graphicx}
\usepackage{hyperref}

\begin{document}

\title{No Free Lunch for Avoiding Clustering Vulnerabilities in Distributed Systems}

\author{Pheerawich Chitnelawong,$^{1}$ Andrei A.\ Klishin,$^{2,3}$ Norman MacKay,$^{1}$
David J.\ Singer,$^{4}$ Greg van Anders$^{1}$}

\address{ $^{1}$Department of Physics, Engineering Physics, and Astronomy,
Queen's University, Kingston ON, K7L~3N6, Canada}
\address{$^{2}$Department of Mechanical Engineering, University of Washington
Seattle, WA 98195, USA}
\address{$^{3}$AI Institute in Dynamic Systems, University of Washington,
Seattle, WA 98195, USA}
\address{$^{4}$Department of Naval Architecture and Marine Engineering
University of Michigan, Ann Arbor, MI 48109, USA}
\ead{gva@queensu.ca (Greg van Anders)}
\vspace{10pt}
\begin{indented}
\item[]August 2023
\end{indented}

\begin{abstract}
Emergent design failures are ubiquitous in complex systems, and 
often arise when system elements cluster.  Approaches to systematically reduce
clustering could improve a design's resilience, but reducing clustering is
difficult if it is driven by collective interactions among design elements.
Here, we use techniques from statistical physics to identify mechanisms by
which spatial clusters of design elements emerge in complex systems modelled
by heterogeneous networks. We find that, in addition to naive,
attraction-driven clustering, heterogeneous networks can exhibit emergent,
repulsion-driven clustering. We draw quantitative connections between our
results on a model system in naval engineering to entropy-driven phenomena in
nanoscale self-assembly, and give a general argument that the clustering
phenomena we observe should arise in many distributed systems. We identify
circumstances under which generic design problems will exhibit trade-offs
between clustering and uncertainty in design objectives, and we present a
framework to identify and quantify trade-offs to manage clustering
vulnerabilities.
\end{abstract}

\maketitle

\section{Introduction}
A key challenge in the design of complex, large-scale systems is managing
emergent
vulnerabilities~\cite{shields2017emergent,Kuhnle2017,Holme2002a,Nesti2018,Duenas-Osorio2009},
especially those driven by clustering.  Examples of cluster-driven
vulnerabilities include failures in communication networks such as the
Internet~\cite{Albert2000,Albert2002,Sun2007}, hot spot leakage in
microprocessors~\cite{Hu2003,Yang2004}, congestion in airline
networks~\cite{Ceria2021,Fageda2016}, and high outfit density in naval
engineering~\cite{Shields2017,Keane2014}. Though clustering vulnerabilities are
inherently local, they typically degrade global, network-wide performance. For
example, mitigating isolated hot spots in microprocessors frequently involves
throttling the performance of the entire device~\cite{Sharma2015,Sharma2016,
Tschanz2002}; or damage to a single shipboard system can cause co-located
systems to fail, inducing failures that cascade through the
ship~\cite{doerry2007,Goodrum2018,HabbenJansen2019,HabbenJansen2020}.
Guidance for mitigating clustering instabilities is therefore critical across
many engineering domains.

Clustering vulnerabilities are ubiquitous in complex system
design~\cite{Watts1998,Holme2002,Holme2002a,Kuhnle2017} because, in systems
comprised of a large number of functional units, units are often arranged to
minimize physical connection distances. Distance-minimization drivers can be
economic, e.g., minimizing material cost~\cite{Whitcomb1998,Watson1998}, or
physical, e.g., minimizing energy loss or
latency~\cite{Cui2007,Benini2002,Jiang2019}, or a combination of economic and
physical~\cite{Benford1967}. Regardless of the driver, connection-distance
minimizing arrangements of design elements induce spatial grouping, and that
grouping produces clustering
vulnerabilities~\cite{HabbenJansen2020a,Brownlow2021}.  Approaches to reduce
clustering vulnerabilities in general therefore need to disrupt the drive for
short connection distances.

Since clustering vulnerabilities arise generically in the optimization of
systems with complex inter-dependencies, there are two obvious strategies to
mitigate such vulnerabilities.  One obvious mitigation strategy is to make ad
hoc modifications to the optimization criteria to counteract the clustering that
produces the vulnerability.  However, vulnerabilities are most difficult to
manage in the system that they are most likely to arise in, e.g., complex
systems described by large, he of connected elements~\cite{shields2017emergent}.
Because emergent vulnerabilities occur unpredictably, actions to mitigate one
vulnerability may induce the emergence of others, which could be more numerous
or severe than the original.  If modifying optimization criteria ad hoc to
mitigate one vulnerability can drive the emergence of others, a second
alternative strategy could be to look beyond strictly optimal solutions in a
systematic way. In situations where economic considerations factor among
optimization criteria considering non-optimal solutions is, in a colloquial
sense, buying a way out of the problem. However, employing expensive,
sub-optimal solutions could be a worthwhile sacrifice if the vulnerabilities
induced by clustering are severe, and if sub-optimal solutions reliably
eliminate clustering.

Here, we show that a complex interplay between how design elements are
physically placed and how their functional connections are placed defeats simple
strategies to mitigate cluster-driven vulnerabilities. We argue that in generic
situations in which there are multiple possible choices of connection routes for
each given placement of functional elements in a fixed spatial region, routing
multiplicity is the dominant driver of element placement when
connection-distance minimization is relaxed. In relaxed distance-minimization
settings, we find that routing multiplicity and connection heterogeneity combine
to drive new, emergent forms of clustering. This emergent clustering defeats
naive ``buy out'' approaches to mitigate clustering vulnerabilities. By
connecting routing and placement multiplicity to analogous concepts of
configurational and conformational entropy in physical systems, we show that it
is possible to avoid clustering by managing a balance between placement and
routing multiplicity, but that this can only be achieved at the expense of high
uncertainty in the original optimization objectives. We give a concrete
illustration of these effects via a detailed analysis of a problem in Naval
Engineering.  Our results show there is no simple, one-size-fits-all approach to
managing clustering vulnerabilities. However, we show that clustering
vulnerabilities can be managed in a context-dependent, case-by-case manner using
a consistent framework.

\section{Results}
\subsection{Clustering in a Model System}

To motivate a general argument on clustering vulnerabilities, it is instructive
to first understand how they arise in a specific example. We use a system from
naval architecture that describes arranging the power system of a naval vessel
inside a ship hull.~\cite{Shields2017} The model describes the placement of
elements of the power system and their interconnections, with a cost associated
with connection length (see Methods~\sref{sec:methods}). This model has two features that are
exemplary of other contexts: (i) optimizing arrangements for short connection
distances drives system elements to cluster in space~\cite{systemphys}; (ii)
because the network is comprised of elements that are power sources and power
sinks, the connectivity of the network is heterogeneous; i.e., it has both high-
and low-connectivity elements.

We study this model by generating candidate arrangements controlled by
a parameter $T$ (see Methods~\sref{sec:methods}) that serves as a tolerance for generating
non-minimal routing distances, and is mathematically equivalent to temperature
(see Methods~\sref{sec:methods}).  $T=0$ indicates no tolerance for non-minimal routing distance
and $T\to\infty$ indicates unlimited tolerance for non-minimal
routing distance. Results below parametrize $T$ relative to a crossover value
$T_c$ that we determine by comparing classes of model results (see Methods~\sref{sec:methods}).

\begin{figure}
	\centering
	\includegraphics[width=\linewidth]{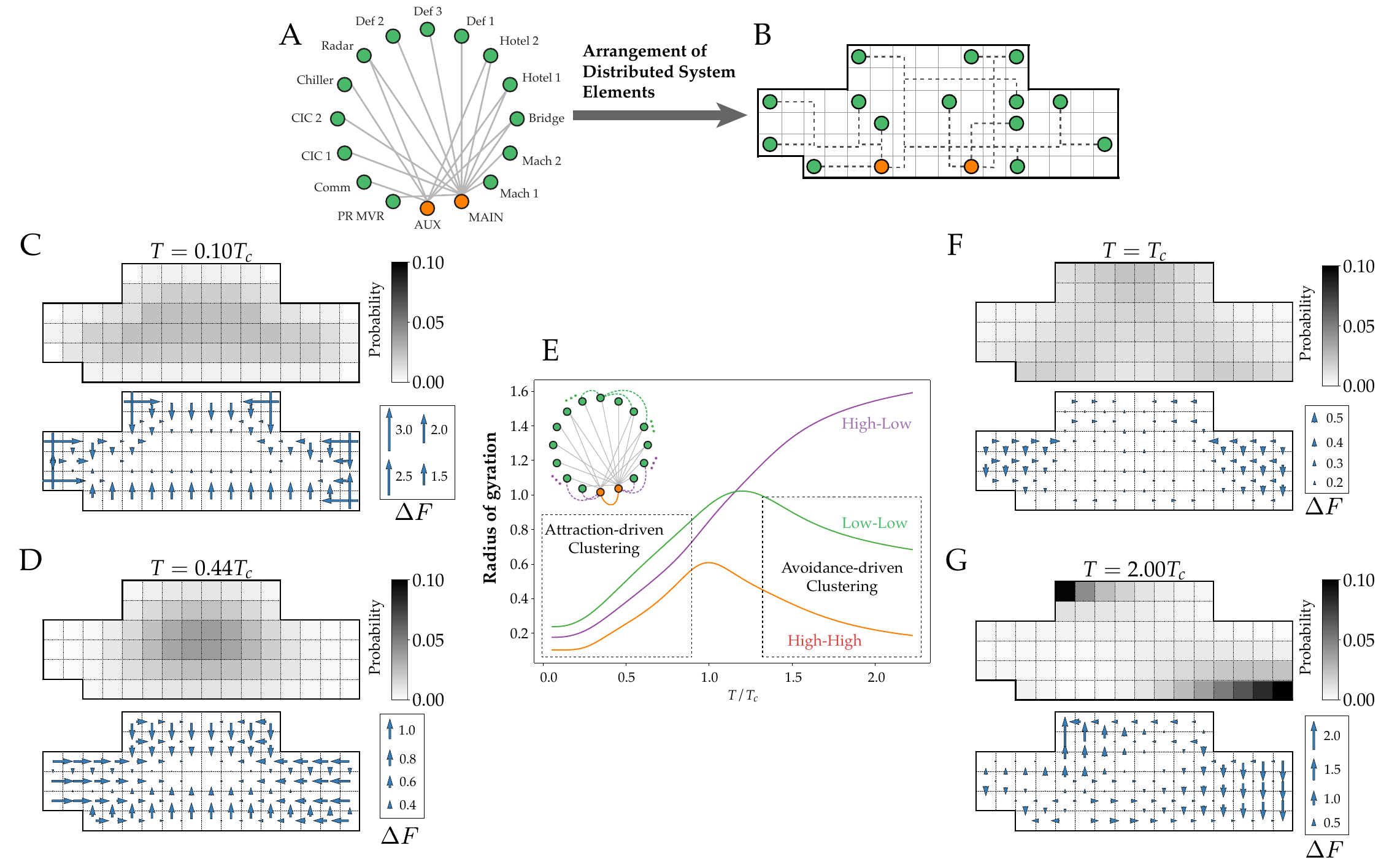}
	\caption{
	\textbf{Avoidance-driven clustering emerges in non-minimal distance routing.}
	Panel~A illustrates design-element connectivity of a shipboard power systems.
	Panel~B illustrates a hypothetical arrangement.
	Panel~E plots radius of gyration ($R_g$) versus T that quantifies correlations
	between elements paired by their degree of connectivity, low-low, high-high, and
	high-low. Low $R_g$ at low T, where designs are dominated by minimal routing
	distance are clustered by attraction. However, low $R_g$ at high T for
	functionally disconnected, low-low and high-high connectivity pairs that
	coincides with high $R_g$ for functionally connected high-low pairs indicates
	repulsion driven clustering. The inset network diagram shows example element
	pairs that $R_g$ is averaged over.
	Panels C, D, F, and G show element localization in the ship hull. Low T
	clustering (C and D) is expected since the objective prioritizes routing cost,
	and the distribution is a single dense region, that is central, and excludes the
	boundary, all consistent with attraction-driven clustering.
	High T clustering (G), however, is multi-modal and peripheral.
	Moderate T (F) corresponds to local peaks in $R_g$ (panel E) and is distributed
	throughout the hull which indicates de-clustering.
	}\label{fig:main2}
\end{figure}

\Fref{fig:main2} reports clustering behaviours in our model system.  We use two types of
measure: global and local. The global measure, radius of gyration, reports
average `pair-wise' (two-point) correlation between all pairs, regardless of the
position of the design element in the ship. The local measure reports an
average one-point correlation over all design elements over different locations
of the ship hull.  We report both global measures of clustering (panel E) and
local measures in the ship hull (panels C, D, F, G) at a range of tolerances for
non-minimal route length.

\subsubsection{Global Measures Show Emergent Clustering of Non-Connected Elements.}
We note that the underlying form of the power system as a set of objects
``tethered'' to one another by functional connections is reminiscent of polymer
systems. So, we borrow from polymer physics and measure global clustering via
the radius of gyration, $R_g$,\cite{Kamide2000} a root-mean-squared distance
between a set of objects (see Methods~\sref{sec:methods}). As expected, at low $T$ we find that
power system elements are tightly clustered, regardless of their degree of
network connectivity, as indicated by low gyration radii.

More surprisingly, however, we find that clustering re-emerges when $T$ is high.
\Fref{fig:main2}E shows that although $R_g$ approaches the size of the space for directly
connected power system elements (all of which are high-low connectivity pairs),
$R_g$ is small for unconnected power system elements (low-low and high-high
connectivity pairs).  This form of clustering is striking for two reasons:
because clustering involves subsets of the elements, and because the elements
that cluster together are ones that don't have direct functional connections.

\subsubsection{Local Measures Show Emergent Clusters are Peripheral.}
To understand the origin of emergent clustering, we analyzed the local
distribution of power system elements throughout the ship hull. We first
establish a baseline for comparison by computing element distributions at
low $T$, where we expect conventional attraction-driven clustering.

\Fref{fig:main2}C shows the distribution of power system elements in
arrangements driven by attraction-driven clustering. The global measure of
clustering at low $T$ in \Fref{fig:main2}E indicated that the system elements form
a cluster with small $R_g$. The local distribution in \Fref{fig:main2}C
shows that the global clustering coincides with arrangements with a near-uniform
distribution throughout the ship hull. 
Note that the exception to uniformity is the depleted region near the boundary.
This behaviour is analogous to the behaviour of polymer solutions confined, for example, within a tube with repulsive walls where the conformational entropy, i.e., internal rearrangements, of the polymer reduce the density of the solution near the walls~\cite{Gennes1979}.
As further confirmation, increased but still relatively low $T$ (from $0.2 T_c$
to $0.4 T_c$) produces clusters with increased $R_g$, which lead to element
distributions with a wider depleted boundary layer. These distributions share
three features: a single dense region, central distribution, and boundary
exclusion.

However, distributions at high $T$ reverse all three features: high $T$
distributions that they are multi-modal, not centred, and do not exclude the
boundary.  On the contrary, \Fref{fig:main2}G shows that high $T$ ($T = 2T_c$) generates
element distributions that exclude the central region; instead they localize on
the boundary in two distinct regions. The existence of these two regions accords
with the global $R_g$ clustering measure: \Fref{fig:main2}E showed that functionally
disconnected elements that had similar degrees of connectivity (low or high)
formed clusters, whereas functionally connected elements had large $R_g$. The
element distribution in \Fref{fig:main2}G suggests the differing clustering of
functionally disconnected elements (low-low and high-high, which cluster) and
functionally connected elements (high-low, which spread) arise because distinct
regions of the element distributions correspond to elements with distinct
degrees of connectivity.

\subsubsection{Emergent, Peripheral Clusters Segregate by Degree of Connectivity.}
\begin{figure}
	\centering
	\includegraphics[width=\linewidth]{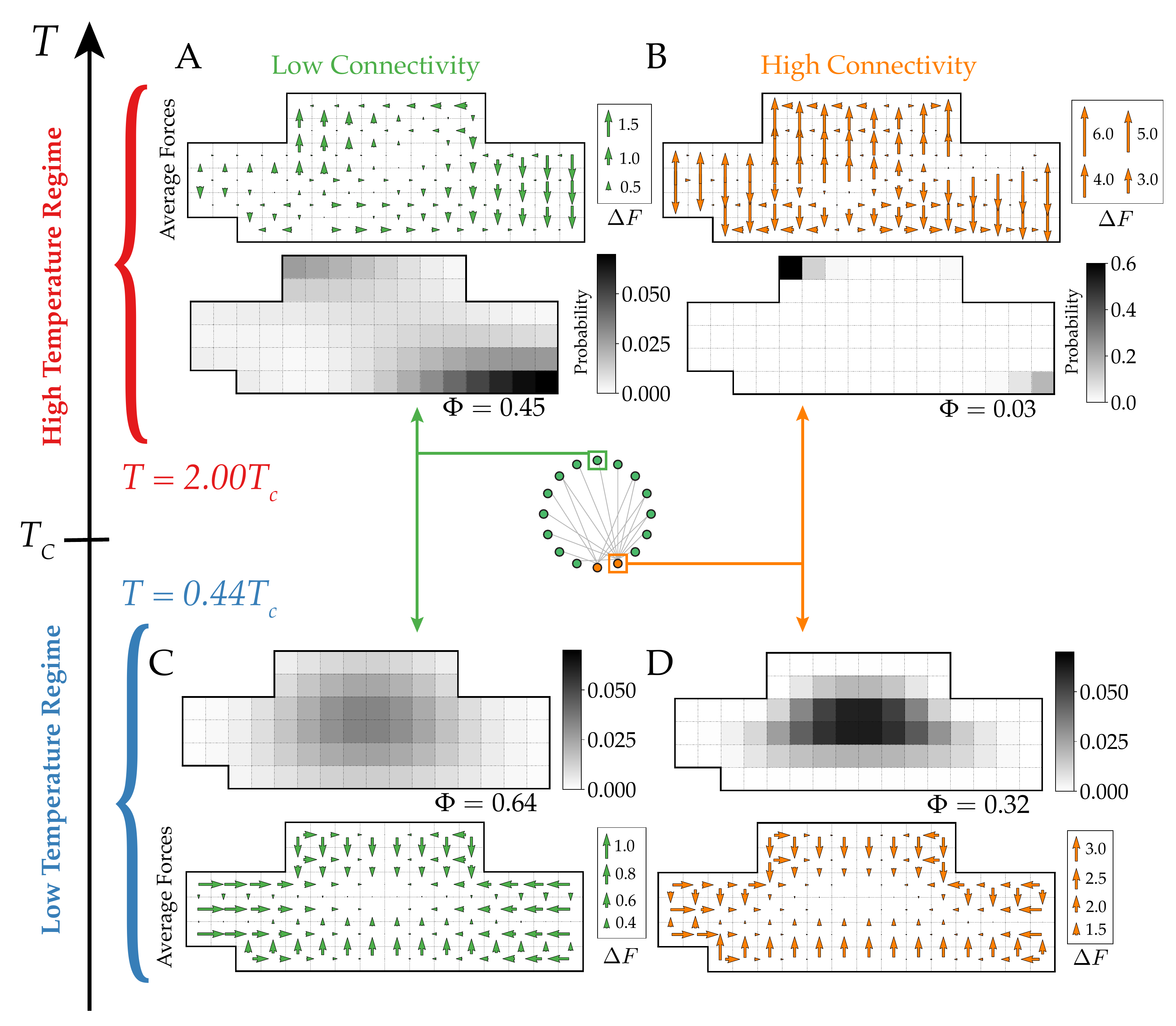}
	\caption{
	\textbf{Emergent clustering segregates design elements by degree of connectivity.}
	Panels A-D show probability distributions and effective forces for
	representative design elements that have low- (AC) or high- (BD) degrees of
	connectivity. Avoidance-driven clustering at high $T$ in low-connectivity
	elements (A) and high-connectivity elements (B) display distinct behavior. Low
	connectivity elements adhere to the bottom right of the hull, and
	high-connectivity units localize near the top left. Effective forces (quiver
	plots) are larger for high-connectivity elements. Attraction-driven clustering
	at low $T$ for low-connectivity (C) and high-connectivity (D) elements shows
	high-connectivity elements are more localized, which is consistent with the
	force measurements.
	}\label{fig:regimes}
\end{figure}

To determine whether cluster separation occurs because elements separate by
degree of connectivity, we separately analyze the distribution of representative
high-connectivity and low-connectivity power system elements in \Fref{fig:regimes}. Computed
distributions indicate that the two concentration areas in the
connectivity-agnostic element distribution in \Fref{fig:main2}G can be associated with
either high- or low- connectivity elements. Panels A and B in \Fref{fig:regimes} indicate
that emergent clusters segregate elements by their degree of connectivity.

For comparison, we computed distributions for the same elements at low $T$.
Panels C and D in \Fref{fig:regimes}
suggest that, while there is considerable overlap between distributions for
high- and low-connectivity elements, high-connectivity distributions are more
concentrated suggesting a ``core-shell'' form of spatial organization.

\subsubsection{Routing Multiplicity Drives Emergent Clustering.}
The existence of clustering at low $T$ is
unsurprising, however the re-emergence of clustering when the preference for
non-minimal routing is relaxed is unexpected. The fact that clusters form in
separate, segregated, peripheral groupings violates the intuition power system
elements should de-localize if the drive for minimal routing is relaxed.

To understand the origin of the unexpected emergent clustering, it is
instructive to extend the analogy with conventional physical systems. The
generating function for arrangements (see Methods~\sref{sec:methods}) can be decomposed into three
sets of contributions: the length-dependent cost of routing connections between
power system elements, the multiplicity of arrangements of power system elements
with fixed route lengths, and the multiplicity of the routing paths for a fixed
element arrangement and route lengths. These factors are analogous to line
tension, configurational, and conformational entropy, respectively, in physical
systems. The identification of these physical analogues gives a direction for
further analysis.

\begin{figure}
	\centering
	\includegraphics[width=0.4\linewidth]{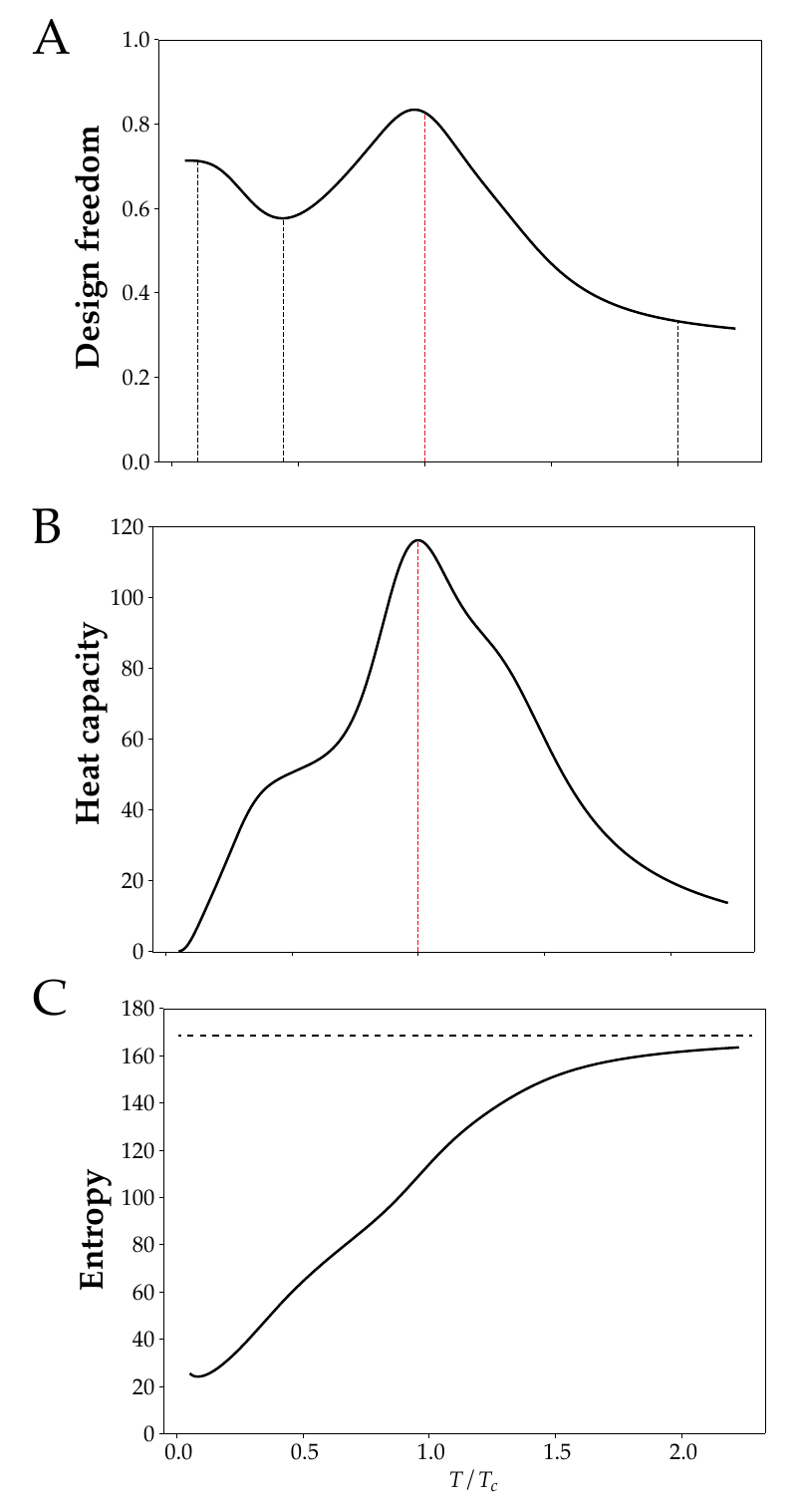}
	\caption{
	\textbf{Declustering coincides with high-variability in the design objective.
	}
	\textbf{Panel~(A)} design freedom (a measure the freedom to place design
	elements, see Methods) vs $T$ peaks at intermediate $T$, indicating
	declustering. \textbf{Panel~(B)}, heat capacity vs $T$, indicates that
	declustering coincides with a peak in heat capacity. However, because heat
	capacity measures fluctuations in routing distance (see Methods), declustering
	coincides with maximal uncertainty in the primary design objective.
	\textbf{Panel~(C)}, total system entropy vs $T$ increases monotonically, as
	expected, indicating that decreasing design freedom at large $T$ with cost
	tolerance occurs because routing multiplicity increases at the expense of
	unit placement multiplicity.
	}\label{fig:main1}
\end{figure}

The analogy between power system element arrangement multiplicity and
configurational entropy suggests quantifying this contribution in terms of the
spatial spread of the element distribution. In other contexts, existence area
(see Methods~\sref{sec:methods}) is used to measure inhomogeneity in distributions that arises in
localization.\cite{filoche2009localization} Here, we use the same mathematical
form to characterize the arrangement multiplicity of power system elements.
\Fref{fig:main1}B shows this form of design freedom, which is a proxy for configurational
entropy, as a function of $T$.  Decreases in design freedom at low $T$ and at
high $T$ are counter-intuitive because they indicate a loss of configurational
entropy.

In thermodynamic systems, entropy conventionally increases monotonically with
temperature, which is typically expressed in terms of a strictly positive heat
capacity, $C_V$. \Fref{fig:main1}A,C show total system entropy and
heat capacity as a function of $T$. \Fref{fig:main1}A shows that entropy
is an increasing function of $T$ as expected, and \Fref{fig:main1}C
shows that the heat capacity is strictly positive. These results indicate that
the reductions in design freedom with increasing $T$ still coincide with
increasing total entropy, but that this total entropy increase occurs because
the conformational entropy associated with the existence of multiple routing
paths between a fixed arrangement of elements overwhelms the configurational
entropy associated with the multiplicity of element arrangement.

Taken together, five factors all suggest emergent clusters that are separate,
segregated, and peripheral is driven by the generation of arrangements that are
dominated by the multiplicity of routing paths for a fixed arrangement of power
system elements: (i) the global clustering $R_g$ (\Fref{fig:main2}E); local element
distributions, both (ii) agnostic of connectivity (\Fref{fig:main2}G) and for
representative (iii) low- (\Fref{fig:regimes}A) and (iv) high-connectivity (\Fref{fig:regimes}B)
elements, and (v) contrasting design freedom and entropy measures.

\subsubsection{Declustering Coincides with Design Objective Variability.}
The above analysis showed that the unexpected re-emergence of clustering at high $T$ was
driven by entropic effects. However, this analysis also revealed a peak in the
heat capacity in \Fref{fig:main1}C at intermediate $T$, and this raises
the possibility of a different scenario to avoid clustering.

In systems of macroscopic numbers of atoms, sharp divergences in heat capacity
signal a phase transition at a corresponding critical temperature. And,
importantly, conventional thermodynamic systems at a critical point typically
develop fractal behaviour, with spatial organization at many different
scales~\cite{goldenfeld}. Multi-scale organization is a possible ``out'' to the
clustering problem, and could be achievable, not at high $T$ where instead we
observed re-emergent clustering, but at intermediate $T$.

The present system has a finite number of elements so the heat capacity cannot
exhibit a sharp divergence. However, despite the lack of a sharp divergence in
heat capacity in the present system, quasi scale-invariant behaviour is possible.
To investigate this we carried out global and local measures of clustering at
$T_c$. \Fref{fig:main2}F shows that indeed power system elements distribute in both
central and peripheral locations. This spread-out distribution approximately
coincides with maximal $R_g$ for non functionally-connected elements (\Fref{fig:main2}E),
and a peak in design freedom (\Fref{fig:regimes}B).

\begin{figure}
	\centering
	\includegraphics[width=\linewidth]{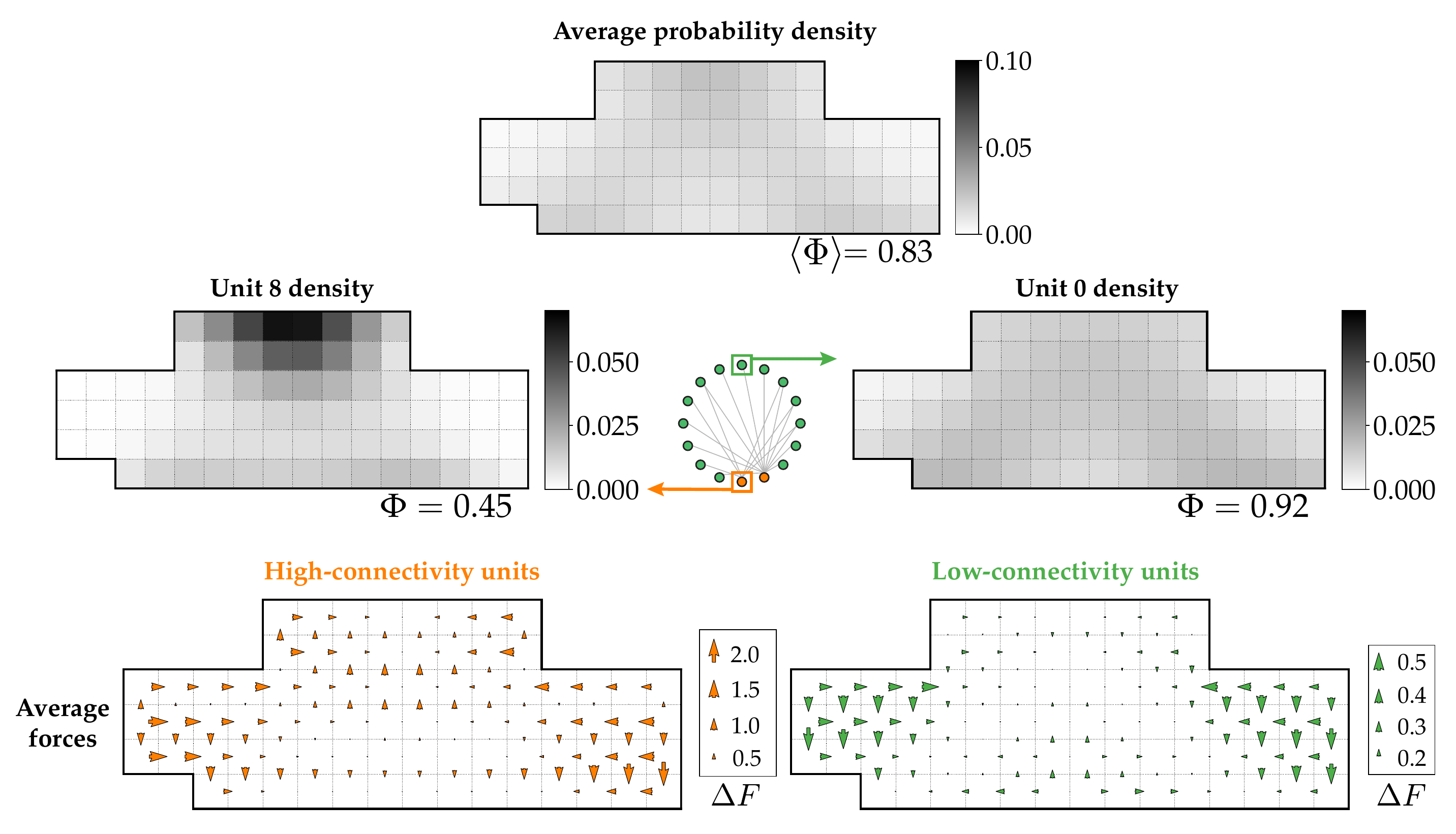}
	\caption{
	\textbf{High variability configurations segregate by connectivity.}
	High variability configurations at intermediate $T$ ($T=T_c$) have distributions
	that differ by connectivity. High-connectivity units (left panels) localize more
	strongly than low-connectivity units (right panels).
	}\label{fig:highvar}
\end{figure}

Furthermore, segregation by connectivity for high- and low- connectivity elements is also found at intermediate temperature as shown in \Fref{fig:highvar}.
However, the clustering behaviour is significantly mitigated with the average design freedom of 0.83 at $T=T_c$. 
This result emphasizes that connectivity-dependent localization is an inherent feature of a heterogeneous network that persists across all temperatures.

Together, the results indicate that in the vicinity of $T_c$ there is a
significant decrease in clustering. However, this decreased clustering comes at
a price. Thermodynamic heat capacity serves as a measure of the size of energy
fluctuations across a set of states.~\cite{Sethna2021} Comparing the present system
with thermodynamic systems, the role of energy is taken by the total routing
cost, and so high heat capacity implies high routing cost uncertainty. This
routing cost uncertainty, however, implies high variability in the main design
objective. This means that in the present system, declustering coincides with
high design objective variability.

\subsection{Clustering by Repulsion-Driven Attraction in General}
The clustering behaviour we observed in the naval architecture model arose from
an interplay between the objective for minimal-length routes and two
multiplicities: arranging power system elements and arranging the connections
between them. These two forms of multiplicity are directly analogous to
configurational and conformational entropy that drive clustering and arrangement
in other systems~\cite{DiamondoidEnt}.
A well-studied example of this is the self-assembly of
tethered nanoparticles.\cite{Zhang2003} In these nanoscale systems, a subtle
interplay between the entropy of the nanoparticle configurations and the
conformations of polymer tethers drives complex, emergent organization,
including the clustering of non-attracting objects.\cite{Marson2015} The
existence of a nanoscale analogue of the behaviour we observed in our naval
architecture model strongly suggests the model behaviours we observed signal a
manifestation of a more general phenomenon.

To see this, consider a generalization of the model arrangement problem we
analyzed above. The general model is a system of $N$ elements to be placed at some
positions $\vec{x}_i$ in a domain $D$, where the subscript labels the element,
and the vector components are coordinates of the position of that element. It is
most concrete to think of the coordinates describing positions in physical
space, however they could also describe positions in the space of element
specifications (e.g., power consumption). We consider a situation in which some
of the objects are functionally connected to one another, and some are not,
which we encode in an adjacency matrix $A_{ij}$, which is one if elements $i$
and $j$ are functionally connected and zero if they are not.

Understanding whether clustering occurs in arranging design elements requires a
means to generate arrangements systematically. Systematically generating
candidate solutions to a problem that makes minimal assumptions about the form
of the solutions is governed the theory of information.\cite{shannon}
Information theory shows~\cite{systemphys} that generating solutions scored by a
design objective, here the length of the routes between elements
$L(\vec{x}_i,\vec{x}_j)$, is described by the generating function
\begin{equation}\label{eqn:partition}
  Z(\beta) = \sum_{\{\vec{x}_i \in D\}} \sum_{\{R(\vec{x}_i,\vec{x}_j)\}}
	e^{-\beta \sum_{i,j} A_{ij} L(\vec{x}_i,\vec{x}_j)}
\end{equation}
where $\beta = 1/T$ is the inverse of the tolerance for non-minimal routing, and
$R(\vec{x}_i,\vec{x}_j)$ is the set of possible connection routes between
$\vec{x}_i$ and $\vec{x}_j$. $Z(\beta)$, which is known as a partition function
in statistical mechanics, is a Laplace transform of the design objective that
generates candidate arrangements at a frequency weighted by a pressure
$\beta$ for minimal routing. I.e., at large $\beta$ (equivalent to $T\to 0$),
\Eref{eqn:partition} generates only arrangements with minimal or near-minimal
routes, and generates routes of increasing $L$ as $\beta\to 0$ (equivalent to
$T\to\infty$).

The form of arrangements that $Z$ generates is determined by how the
multiplicity of options for routing a connection grows with the connection
length. To see this, take the cardinality of the set of routes $R$
between $\vec{x}_i$ and $\vec{x}_j$ as $\Omega(R(\vec{x}_i,\vec{x}_j))$ which
gives the generating function as
\begin{equation}\label{eqn:gen_func}
  Z(\beta) = \sum_{\{\vec{x}_i \in D\}}
  e^{-\beta \sum_{i,j} A_{ij} \left(L(\vec{x}_i,\vec{x}_j)-\frac{1}{\beta}
  \ln\Omega(R(\vec{x}_i,\vec{x}_j))\right)}
\end{equation}
The factor in the exponent can be written as an effective distance
\begin{equation}\label{eqn:exponent}
  \Delta(\vec{x}_i,\vec{x}_j) =
  L(\vec{x}_i,\vec{x}_j)-T\ln\Omega(R(\vec{x}_i,\vec{x}_j))
\end{equation}
which expresses that routing multiplicity $\Omega$ counteracts the drive for
minimal length $L$ with a strength that is determined by the threshold for
non-minimal routing $T$. Notably, $\ln\Omega$ is the Boltzmann entropy in
statistical mechanics, the same physical property that drives clustering
observed at the nanoscale.\cite{Zhang2003,Marson2015} This quantitatively
connects known nanoscale clustering mechanisms to arrangement clustering. 
In routing problems where $\ln\Omega(R(\vec{x}_i,\vec{x}_j))$ grows sufficiently
fast, e.g.\ combinatorially, there will be a threshold $T$ that induces
$\Delta(\vec{x}_i,\vec{x}_j) < 0$ via entropic repulsion.

In microscopic systems, entropic repulsion generates clustering at
inhomogeneities either generated by symmetry breaking~\cite{Zhang2003,Marson2015} or at pre-existing inhomogeneities at
boundaries.~\cite{dinsmore1996entropic} Bounded, inhomogeneous domains, which
induce sites of microscale clustering,\cite{dinsmore1996entropic} are also
generic in distributed systems. In our model system, it was precisely at
boundary inhomogeneities clustering emerged, and this phenomenon should be
generic.

Our analysis indicates that clustering occurs generically, and is driven by one
of two mechanisms. Design elements can cluster to minimize connection
distance, i.e.\ by minimizing $|\Delta|$ for $\Delta>0$ in \Eref{eqn:exponent}. Or, elements can cluster emergently by entropic repulsion,
i.e.\ by maximizing $|\Delta|$ for $\Delta<0$ in \Eref{eqn:exponent}. Since
cluster occurs for both $\Delta>0$ and $\Delta<0$, the only way to avoid
clustering is if $\Delta\approx 0$ for separation distances that are larger than
the characteristic $R_g$ of attraction-driven clusters and smaller than the
separation distance of boundary inhomogeneities where repulsion-driven clusters
localize. However, when $\Delta\approx 0$, the effects of minimal routing length
and routing multiplicity counteract one another. In this regime, the system is
driven by the configurational entropy of the arrangement of the elements, with
the result being large variability in $L$. This means that avoiding clustering
only occurs at the expense of high variability in the design objective.

\section{Discussion}
Technical systems that manifest the adage that there is ``no free lunch'' have
been identified in studies of search and optimization algorithms, where a series
of theorems about algorithm performance averaged over all problem
instances~\cite{Wolpert1995,Wolpert1997}. It was shown it is impossible to
select an \emph{a priori} better algorithm unless one knows particular features
of search or optimization landscapes that can be exploited, such as smoothness,
differentiability, or convexity.  Many similar no free lunch results have been
proven for other problem domains~\cite{Adam2019}, including community detection
on networks where no algorithm is optimal for finding communities across all
types of networks~\cite{Peel2017}. Real-world networks are, however, very
rarely unstructured, and any realistic structure can be exploited to improve the
algorithm, leading to a ``cheap lunch'' effect~\cite{Peixoto2022}.

While search and optimization results rely on a single optimality criterion
to deny the free lunch, our results suggest the appearance of distinct
mechanisms of non-optimality. Though we set out to solve a distributed system
arrangement problem in such a way as to disrupt clustering, we discovered that
disrupting clustering can only be achieved at the cost of design uncertainty.
If we had \emph{a priori} valuation of clustering avoidance over design
uncertainty (or vice versa), we would readily prefer a particular $T$ regime
over others.  This behaviour is, however, precisely the exploitation of prior
knowledge that allows one to circumvent the no free lunch theorems in other
domains. While our system shows two distinct failure modes of the design
process, clustering and uncertainty, it is useful to draw connections to the
clustering effects that appear in other contexts.

The clustering effects we observed in this paper depends on the close interplay of
network topology and spatial constraints~\cite{aaa}, but is distinct from the
effects reported in many spatial network studies. Space-first studies usually
first fix the spatial locations of the nodes and either study the empirical
topology of the links, or propose a distance-based model of link
probability~\cite{Barthelemy2011}.  Depending on the model parameters, the
resulting spatial networks can manifest different global topological features,
such as the small-world effect, topological clustering, or assortative mixing by
node degree~\cite{Makarov2018,Boguna2020}. In contrast, network-first studies
start with a network topology and map nodes to coordinates into a
high-dimensional Euclidean or hyperbolic space to study community structure,
link prediction, and network navigability~\cite{Krioukov2010,Zhang2021}.  While
those approaches illuminate many spatio-topological features of large real-world
networks, neither is able to describe embedding a fixed network into a
prescribed low-dimensional space with a fixed complex boundary.

Directly embedding networks into a low-dimensional space (2D or 3D) is important
for two reasons: to provide manufacturing prescriptions as in our problem, or to
visualize networks on paper or screens.  When both nodes and edges have finite
size and can't overlap, the resulting spatial embeddings show different regimes
of complex structure and mechanical response, mirroring the constraints on
neuronal connections in mammalian brains induced by the
space~\cite{Dehmamy2018}.  When the space constrains our visualizations of
networks to a 2D screen or page, the ambiguity of resulting pictures aids in
exploratory data analysis~\cite{Venturini2021}.  The visual structure of such
network layouts is often created through force-directed layout algorithms and
thus naturally highlights the community structure of the
network~\cite{Zhang2021}. The network used in the present paper has a strongly
disassortative and bipartite structure as connections only exist between
low-degree and high-degree nodes, thus leading to the segregation of nodes by
degree in the high $T$ regime (\Fref{fig:regimes}).

In this paper we explored the problem of embedding fixed networks into a
low-dimensional space with a fixed boundary that has received little theoretical
attention despite its practical importance for systems from microprocessors to
airlines~\cite{Hamann2007,Lavandier2021}, where close spatial proximity of nodes
leads to vulnerability. We show how the combinatorial space of possible routings
necessarily gives rise to either attractive or repulsive clustering, or high
design variability.  With no \emph{a priori} preference between those regimes,
there is no problem-solving advantage at any value of T, and thus no free lunch.
While the \emph{existence} of spatial clusters is a necessity in low-T and
high-T regimes, the \emph{structure} of those clusters can be affected by the
topological features of networks, such as broader degree distributions and
assortative mixing. The clustering behaviour can also be different in systems
with a much smaller number of routings $\ln\Omega(L)\lesssim L$, or systems that
allow for more expensive, non-minimal routing paths.

\begin{figure}
	\centering
	\includegraphics[width=\linewidth]{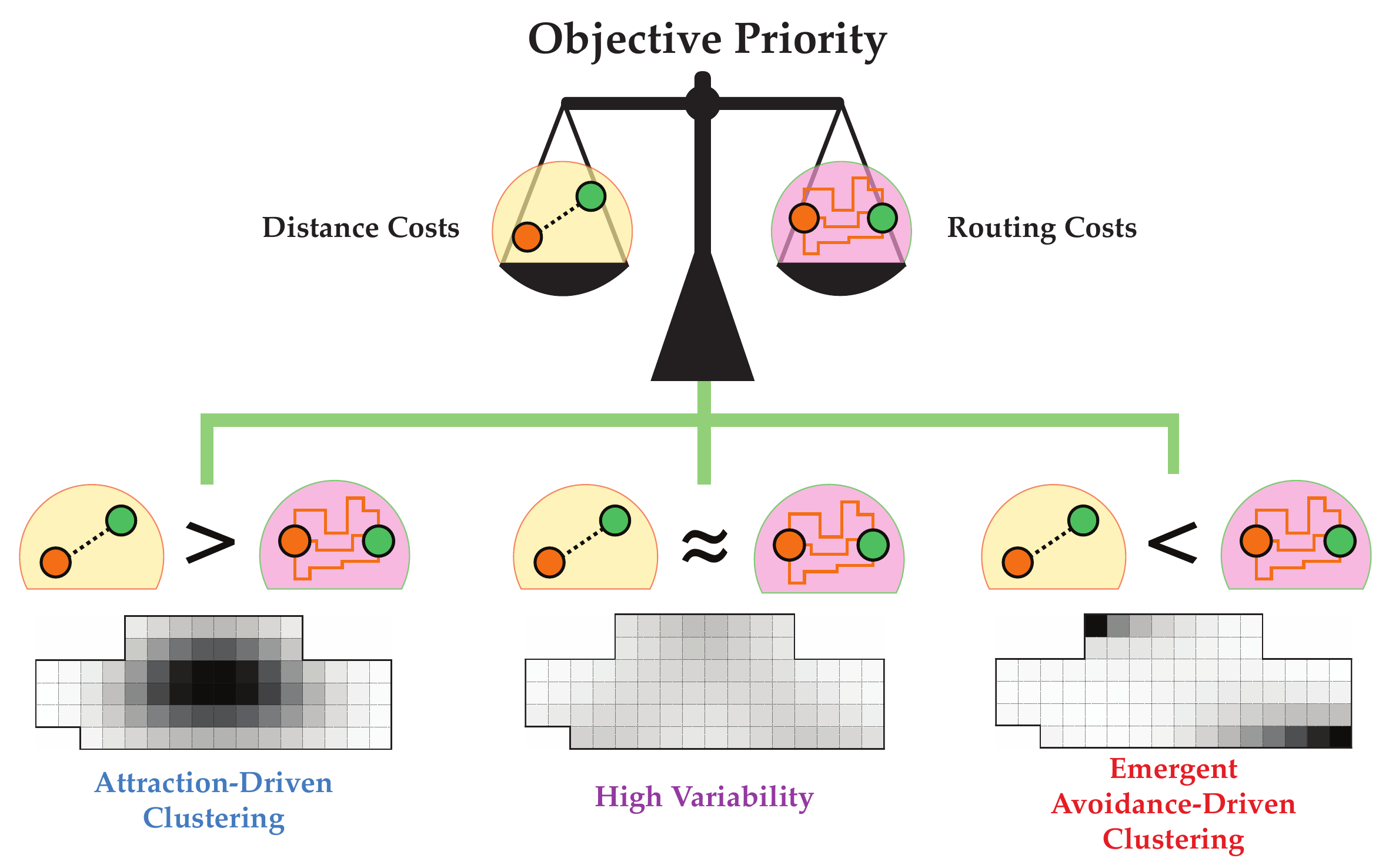}
	\caption{
	\textbf{Trade-offs between multiple forms of clustering and objective variability
	imply there is no free lunch to disrupt clustering.} The attraction-driven
	clustering is an effect of the distance cost degrees of freedom, while the
	emergent avoidance-driven clustering is of the domination of the routing cost
	degrees of freedom.  The configurations where the competing degrees of freedom
	are balanced produce high variability.  Adopting high variability solutions
	circumvents the emergent clustering altogether.
	}\label{fig:concept}
\end{figure}

We have shown that there is a complex mechanism involving the competing degrees
of freedom and how element connectivity is arranged, and this competition of
degrees of freedom drives emergent clustering.  Furthermore, we have shown that
the behaviour of clustering is dependent on the degree of connectivity of the
design elements.  This complex mechanism can be avoided altogether by adopting
high variability configurations.  Using the high variability strategy, while we
lose the certainties in arrangement, the clustering vulnerabilities can be
mitigated.  Nevertheless, strategies that satisfy only one type of degrees of
freedom by either adopting solutions at low $T$ or high $T$ also
lead to emergent clustering, and hence creates clustering vulnerabilities.
These trade offs in degrees of freedom shown in \Fref{fig:concept}
further emphasize that there is no free lunch in avoiding emergent clustering.
If $\ln\Omega(L) \lesssim L$, then it is conceivable to buy out. This could open
up strategies in situations where there are routing constraints.

\newpage
\section{Methods}\label{sec:methods}

\subsection{Model}
For the naval architecture model system, we chose the power system elements and connectivity from Ref.~\cite{Shields2017} shown in Panels I and II in \Fref{fig:main2}.
The power sources, MAIN and AUX, are the design elements with high connectivity.
The low-connectivity elements are not logically connected to one another.
The chosen network is a practical arrangement of possible connectivity configurations.
With this connectivity, we construct a thermodynamic model using Systems Physics~\cite{systemphys}.

For the design geometry of the model, we reproduce the geometry from Ref.~\cite{Shields2017} with twice the resolution in each axis, resulting in more available spaces for design elements to occupied.
The geometry is chosen such that it reflects a practical requirement in naval ship design from Ref.~\cite{Shields2017}

To investigate the nature of clustering vulnerabilities, we created a computational method described in the following section to compute a vital statistical property known as partition function for the model distributed system.
The model Naval Engineering system contains 16 design elements of which 14 elements are low connectivity (degree $k \leq 2$) and 2 elements are high connectivity ($k > 2$).
The logical connections between the elements are shown in panel (A) in \Fref{fig:main1}.
To obtain a design solution/configuration, the elements are placed in the ship geometry.
An example of a design solution for the model distributed system is shown in panel (II) in \Fref{fig:main2}.

\subsection{Route Prioritization \& Temperature}
The two types of degrees of freedom, placement and routing, are encoded into the system via the objective function:
\begin{equation}\label{eqn:objt}
	\Delta(\vec{x}_i, \vec{x}_j; T) = C L(\vec{x}_i, \vec{x}_j) - \frac{1}{T}\ln\Omega(R(\vec{x}_i, \vec{x}_j))
\end{equation}
where $C$ is the cost scaling constant set to one in this study, $L$ is the Manhattan distance between two points, and $\Omega(R(\vec{x}_i, \vec{x}_j))$ is the function which encodes the number of shortest path between two spatial positions.
$\vec{x}_i$ denotes the position of the design element in the design geometry.
On one hand, the placement degree of freedom is shown as Manhattan distance which requires more resources as the distance between any two elements increases.
On the other hand, the routing is a competing degree of freedom where the number of shortest routings increases with the distance, acting against the placement degree of freedom.
To examine the trade offs between the two degrees of freedom, we parametrize the priority for short, inexpensive connections through $T$ dampening the Manhattan distance term.
This $T$ is mathematically equivalent to temperature.

Using the objective function from \Eref{eqn:objt}, we calculate the partition function using methods from systems physics~\cite{aaa} shown as:
\begin{equation}
	Z = \sum_\alpha e^{-\frac{1}{T}\sum_{i,j} A_{ij} \Delta(\alpha)}
\end{equation}
where $\alpha$ denotes a \emph{design solution} in a combinatorically large set of candidate designs ${\alpha}$.
A design solution $\alpha$ contains a configuration of the placements of the design elements in the model system.
$A_{ij}$ is the adjacency matrix of element connectivity represented by the network diagram in Panel I in \Fref{fig:main2}.
We compute the partition function, using the computational package \textit{Lachesis}~\cite{Chitnelawong2023}, from design requirements using the tensor network construction demonstrated in Ref.~\cite{aaa}.

With the requirements of the model system, we investigate the thermodynamic behaviour of each type of connectivity: high and low.
The design elements that are logically connected should directly affect clustering since the objective function relies on the possible routing path and distance.
We allow the system to vary the cost tolerance, $T$, where the resources are controlled such that we can examine clustering at various conditions.
At low $T$, the design elements are penalized by being at a distance to one another, since the allowed resources are lower.
While at high $T$, although the resources are more available, the cost of routing can be a key factor in penalizing certain expensive configurations.
Thus, the clustering vulnerabilities may emerge from temperature conditions at both high and low $T$.

\subsection{Critical Temperature Determination}
In a system with the thermodynamic limit, one common indicator of a phase transition is the divergence of the isochoric heat capacity~\cite{goldenfeld}.
A phase transition indicates different regimes of thermodynamic behaviour, and hence, we utilize the temperature at which a phase transition occurs to define the boundary between two distinct regimes of behaviour.
However, in a finite-sized system, the divergence manifests in the form of a maximum~\cite{Chomaz1999}.
To determine the critical temperature $T_c$ in our system, we identify the temperature at which the heat capacity is maximum as shown in \Fref{fig:main1}C.

\subsection{Connectivity-Based Clustering: Radius of Gyration}
For the effects of clustering from connections from any pair of elements, we use the order parameter radius of gyration.
The radius of gyration quantifies the likelihood of a pair of chosen elements to be close to one another, and is defined by:
\begin{equation}\label{eqn:r_g}
R_g = {[\langle p(\vec{x}_i, \vec{x}_j)({(\Delta_{ij} x)}^2 + {(\Delta_{ij} y)}^2) \rangle]}^{1/2}
\end{equation}
where $p(\vec{x}_i, \vec{x}_j)$ denotes the two-point correlation function between two units $i$ and $j$, and the difference $\Delta$ is calculated as the distance between two points.
We use an average, denoted by $\langle\cdots\rangle$, of correlation and distance between all possible pairs of elements to determine the radius of gyration.
Hence, the radius of gyration represents an average clustering between all pairs of elements.

\subsubsection*{Connectivity-Agnostic Clustering: Design Freedom.}
Measuring clustering without reference to element connectivity is akin to
measuring how elements localize in a space. Emergent localization is a
well-studied phenomenon in physics~\cite{thouless1974disordered}, and we follow
Ref.~\cite{aaa} and use a normalized version of the existence area defined by
\begin{equation}
	\Phi = \frac{1}{Y_0} \frac{{\bigg(\sum\limits_{\vec{x}}p(\vec{x})\bigg)}^2}{\sum\limits_{\vec{x}}{p(\vec{x})}^2}
\end{equation}
where the normalization $Y_0$ represents the number of available cells in the
system, and $p(\vec{x})$ is a distribution of unit arrangements. In this form,
$\Phi$ is bounded above by one when $p(\vec{x})$ is uniform, and decreases
monotonically to $1/Y_0$ as $p$ becomes more localized.  Because $\Phi$ describes
the effective fraction of the total area free to unit placement, we refer to
this as design freedom.

\subsection{Design Objective Uncertainty}
For the heat capacity, we find that it has a maximum at a temperature at which we define as a critical temperature of the model system.
A divergence in heat capacity is an indicator of a phase transition in magnetic or other conventional thermodynamic systems~\cite{goldenfeld}.
The present system is finite-sized, so the heat capacity cannot diverge.
A maximum in heat capacity indicates that there are distinct regimes of behaviour.
In our case, the maximum in heat capacity suggests that there may be different behaviour in the temperature regimes separated by the maximum.
Consequently, we define the temperature $T_c$ at which the maximum occurs as the critical temperature in the system.
Panel C in \Fref{fig:main1} shows a finite maximum of heat capacity of the model system which we use to define the critical temperature.

\section*{Acknowledgements}
We acknowledge the support of the Natural Sciences and Engineering Research
Council of Canada (NSERC) grants RGPIN--2019--05655 and DGECR--2019--00469. This
work was supported by the U.S. Office of Naval Research Grant Nos.\
N00014--17--1--2491 and N00014--15--1--2752.

\clearpage
\section*{References}

\providecommand{\newblock}{}

\end{document}